\documentclass[prb,showpacs,twocolumn,amsmath,amssymb]{revtex4}
\usepackage{epsfig}
\usepackage{epsf}
\usepackage{amssymb, amsmath}
\usepackage{graphicx}
\usepackage{dcolumn}
\usepackage{bm}
\usepackage{appendix}

\newcommand{\up}{\uparrow}
\newcommand{\dn}{\downarrow}
\newcommand{\vk}{{\bf k}}
\newcommand{\bfk}{{\bf k}}

\newcommand{\vq}{{\bf q}}

\usepackage{color}
\usepackage{ulem}
\begin{document}

\title{
Andreev  spectroscopy  of Majorana states 
in topological  superconductors with multipocket Fermi surfaces}
\author{Ana C. Silva$^{1}$,  Miguel A. N. Ara\'ujo$^{1,2,3}$
and Pedro D. Sacramento$^{1,3}$}
 

\affiliation{$^1$ CFIF and CeFEMA, Instituto Superior
T\'ecnico, Universidade de Lisboa, Av. Rovisco Pais, 1049-001 Lisboa, Portugal}
\affiliation{$^2$  Departamento de F\'{\i}sica,  Universidade de \'Evora, P-7000-671, \'Evora, Portugal}
\affiliation{$^3$ Beijing Computational Science Research Center, Beijing 100089, China
}

\begin{abstract}
The topological properties of a multiband
topological superconductor
in two dimensions are studied,
when the latter is obtained by introducing electron pairing in an 
otherwise topological insulator.  
The type of pairing, doping and Fermi surface topology play an essential
role.  
Considering the Andreev reflection problem,
we use a previously developed
quantum waveguide theory for multiorbital systems 
and find that, when the Fermi surface has several pockets,
this theory retrieves  the correct number of Majorana fermion states 
as predicted by the topological index. 
By varying band structure parameters, the Fermi surface topology of the normal phase
can be made to change, whereby the number of Majorana modes also varies.
 We calculate  the effect of such
transitions on the Andreev differential conductance.
\end{abstract}


\maketitle

\section{ Introduction}

Recent interest in non-trivial topological properties of 
insulators\cite{hasankane}
has spurred intensive  research on band  models displaying  non-trivial topology\cite{pequim-1}.
In the case of superconductivity, non-trivial topology is associated
with the emergence of  zero energy excitations  which are their own
antiparticles, or Majorana fermions (MFs). 
Condensed matter therefore provides the ground for the realization of such long sought exotic particles
\cite{QiZhang,alicea}.

Early theoretical models of two-dimensional topological superconductivity
 consider $p+ip$  pairing in an otherwise trivial band.  
In such a framework, 
the emergence of MFs
in spinless fermion models was shown to be related to topology\cite{alicea}.
Non-abelian MF's  were shown to arise
at vortex cores in a model of spinfull fermions with spin  triplet $p+ip$  pairing 
and where  the two spin components effectively   decouple\cite{ivanov}. 
It is this concept of MF that we shall deal with, in this work.
Besides their interest from the point of view of fundamental physics, the MF's
 non-Abelian exchange statistics is their  most remarkable
property and  suitable for quantum computation applications \cite{nayak}. 
It was early realized that an effective  $p+ip$ superconductor arises  
from single band Dirac electrons complemented with s-wave pairing\cite{FuKane}.
A promising venue for the construction of topological superconductors
is the proximity coupling of a topological insulator to a superconductor
\cite{engeneering,nature}.
The possibility of engeneering topological superconductors 
using the surface of a three-dimensional TI or a two-dimensional semiconductor
in proximity to a s-wave superconductor was discussed\cite{engeneering}
but multiorbital effects were not considered. 
Non-conventional
superconductivity and Andreev reflection were also studied\cite{lindernagaosa}. 
We note that topological materials are necessarily multiorbital. 
The consideration of  $p+ip$  (or other) pairing 
in a single band model is therefore a simplification.

More recently, there has been  interest in multiorbital systems.
The orbital degree of freedom (pseudospin)  
is present in a lattice symmetry transformation, namely, 
that of lattice inversion, and the characterization 
of pairing term as having  even/odd parity under inversion. 
Such investigations have been motivated 
by the three-dimensional topological superconductor
Cu$_x$Bi$_2$Se$_3$, which has two orbitals per lattice cell.
Some theorems on  the topological indices 
to be expected for various superconductor models, by taking  their symmetries into account,
have recently been established. In many cases, the topology of the Fermi surface
itself is important.
For instance, 
under the assumption of 
lattice inversion symmetry, time-reversal invariance (TRI)
and odd parity pairing,
three-dimensional superconductors are  topological if they possess 
an odd number of Fermi surface pockets\cite{Fu1,Fu2}.
Models of two-dimensional superconductors with a pseudospin
degree of freedom have been proposed recently, concentrating on the case
of nodeless odd parity pairing in TRI superconductors. In this case
non-trivial topology requires spin-orbit couplings non-diagonal
in the pseudospin channel\cite{viola1,viola2}. 

we introduce below a model for a superconductor which has 
two orbitals per lattice site. 
For the case of diagonal (in pseudospin space) pairing,
 we find that  the topological properties
of the kinetic energy have no influence on  those of the superconductor.  
The latter are determined, instead, 
by the pairing symmetry and the Fermi surface (FS) topology.  
We also study the Andreev problem
at the normal/superconductor (N/S) boundary, 
assuming the normal metal to be single band, 
using the previously established
quantum waveguide theory (QWT).
QWT was originally developed to address the Andreev reflection problem in 
heavy-fermion superconductors\cite{euheavy,euNeto} and, later on,
iron pnictide superconductors \cite{waveguide,waveguide2,devyatov1,devyatov2}.
It can be applied when the N/S interface is parallel to one the primitive vectors of the superconductor and normal metal.
For other interface angles, other approaches have to be employed. QWT also ignores the detailed microscopic behavior of 
hopping parameters at the interface.
Refinements of this theory which consider fully microscopic  behavior of  hopping parameters 
 can be found in Refs \cite{referee1,referee3}.

While the 
Blonder-Tinkham-Kaplwjck (BTK) theory\cite{btk} predicts a zero energy Andreev bound state (ABS)
for each FS pocket,  QWT accounts for the
quantum  interference effects between different FS pockets. 
As we shall see below, 
the ABS predicted by  QWT correctly reconciles the Andreev problem with the topological index.

\section{Multiorbital superconductor}
\subsection{Pairing in a 
two-dimensional 
TI}
\label{theorem}

The simplest version of superconducting pairing   is that  
where  electrons with opposite spins 
are related by time-reversal in the normal phase.
The model includes at least two orbitals per lattice site (pseudo-spin)
in addition to the spin degree of freedom. 
In the simplest version,   
the spin $\up$ electrons have  kinetic energy 
$\Xi_\up(\bfk)
$
which is related to that of the spin $\dn$  electrons by  time-reversal,
$\Xi_\dn(\bfk)=\Xi_\up^*(-\bfk)$. 
If $\Xi_\up$ has an odd Chern number then the normal system serves 
as a model for a TI in two dimensions\cite{hasankane}. 
In the case of two orbitals per lattice cell, we can write
\begin{equation}
\Xi_\up(\bfk) =\boldsymbol h(\bfk)\cdot\boldsymbol\tau + h_0(\bfk)\tau_0
\label{htau}
\end{equation}
where the Pauli matrices
$\tau_{i=1,2,3}$ act on orbital space (pseudo-spin) and $\tau_0$ denotes the identity matrix.
We shall later use Pauli matrices $\sigma_i$, $t_i$ and  $r_i$ operating
on spin, particle-hole and Bogolubov-de Gennes amplitude (u v) spaces, 
respectively.
  
The Bogolubov-de Gennes (BdG) matrix in the particle-hole basis
$(\hat c_\up \hat c_\dn \hat c_\up^\dagger \hat c_\dn^\dagger)$,
where the field operators $\hat c_\sigma$ include the pseudo-spin
degrees of freedom,
takes the form:
\begin{eqnarray}
{\cal H}=
\left( \begin{array}{cc}
\hat\Xi & \hat \Delta\\
\hat\Delta^\dagger & - \hat\Xi^T
 \end{array}\right)\,, 
\label{XiDelta}
\end{eqnarray}
where 
$\hat\Xi = {\rm diag}(\Xi_\up,\Xi_\dn)$.
Keeping spin $s_z$ as a good quantum number, the pairing 
matrix $\hat\Delta$ can have
singlet $\psi(\bfk)i\sigma_2$
and triplet $d_z(\bfk)\sigma_1$ components. The form of
the the pairing matrix in pseudospin space remains to be chosen.
We shall write 
\begin{eqnarray}
\hat\Delta(\bfk)=\left(\Delta_s\psi(\bfk)i\sigma_2
+ \Delta_td_z(\bfk)\sigma_1\right)\tau_j\,,
\label{pairing}
\end{eqnarray}
where the Pauli matrix $\tau_j$ remains to be specified.
Then $\cal H$  splits into two BdG matrices for 
$(\hat c_\up \hat c_\dn^\dagger)$ and $(\hat c_\dn c_\up^\dagger)$ spaces. 
The former reads
\begin{eqnarray}
H
&=& \left( \boldsymbol h(\bfk)\cdot\boldsymbol\tau + h_0(\bfk)\tau_0\right) r_3
\nonumber\\
 &+& {\rm Re}\left[\Delta(\bfk)\right]\tau_jr_1 
 - {\rm Im}\left[\Delta(\bfk)\right]\tau_jr_2
 \,,
\label{Huv1}
\end{eqnarray}

We next study the topological indices of (\ref{Huv1}). 
We first consider diagonal pairing
in orbital space,    $\tau_j=\tau_0$. 
In this case the eigenfunctions 
of (\ref{Huv1}) can be written as a direct product 
\begin{eqnarray}
\left( \begin{array}{c}  (u_\up) \\ (v_\dn) \end{array}\right)  =
\left( \begin{array}{c}  u  \\ v \end{array} \right) 
\otimes
\left( \begin{array}{c} \alpha \\ \beta \end{array}\right)
\label{uavb}
\end{eqnarray}
Wave function normalization reads $|u|^2 + |v|^2 = |\alpha|^2 + |\beta|^2 = 1$.
The amplitudes $(\alpha, \beta)$ diagonalize the kinetic energy
\begin{eqnarray}
\left[ \boldsymbol h(\bfk)\cdot\boldsymbol\tau + h_0(\bfk)\tau_0  \right] 
\left( \begin{array}{c} \alpha \\ \beta \end{array}\right) 
= \xi(\bfk)  \left( \begin{array}{c} \alpha \\ \beta \end{array}\right)\,,\label{bandastau}
\end{eqnarray}
and the BdG amplitudes $(u, v)$ obey
\begin{eqnarray}
\boldsymbol h'\cdot\boldsymbol r 
\left( \begin{array}{c} u \\ v \end{array}\right) =  E(\bfk)
\left( \begin{array}{c} u \\ v \end{array}\right)\,.\label{bandasBdG}
\end{eqnarray}
Where we have defined the vector $\boldsymbol h'$ components as
 $h_x'-ih_y' = \Delta(\bfk)$ and $h_z'=\xi(\bfk)$.
The two normal state bands are 
$\xi(\bfk) = \pm |\boldsymbol  h(\bfk)| + h_0(\bfk)\tau_0$.
The Berry connection obtained from (\ref{uavb}) is 
\begin{eqnarray}
{\cal A} &=&
i \left(  (u_\up)^\dagger\ (v_\dn)^\dagger  \right )\frac{\partial}{\partial\bfk}
\left( \begin{array}{c}  (u_\up) \\ (v_\dn) \end{array}\right) \nonumber\\
&=& i(u^*\ v^* ) \frac{\partial}{\partial\bfk}
\left( \begin{array}{c} u \\ v \end{array}\right) + 
i(\alpha^*\ \beta^* ) \frac{\partial}{\partial\bfk}
\left( \begin{array}{c} \alpha \\ \beta \end{array}\right)\nonumber\\
&\equiv&
\boldsymbol a' + \boldsymbol a
\end{eqnarray}
where  $\boldsymbol a'$ and $\boldsymbol a$
denote Berry connections
associated with $(u\ v)$ and $(\alpha\ \beta)$, respectively.
The total Chern number, $C$, is given by 
the line integral around the BZ,
\begin{eqnarray}
C&=&\sum \oint {\cal A}\cdot\delta\bfk\label{Adef}\\ 
&=& 
\sum \oint \boldsymbol a'\cdot\delta\bfk + \sum_\pm \oint \boldsymbol a\cdot\delta\bfk
\label{Cpartial}
\end{eqnarray}
where the summation in (\ref{Adef}) is over the two negative BdG bands
$ E_\pm(\bfk) = -\sqrt{\xi_\pm^2(\bfk) + |\Delta(\bfk)|^2 }$
of the eigenproblem  (\ref{bandasBdG}). The  summation in 
the second term of equation 
(\ref{Cpartial}) is the Chern number of the two normal bands $\xi(\bfk)$ 
and therefore equates to zero.
Hence we find that {\it the topology of $\Xi_\up$ does not 
contribute to $C$} because of cancelation in the sum over
the normal state bands.
It remains to analyse the first term in equation (\ref{Cpartial}).
Using Stokes theorem, the line integral of $ \boldsymbol a'$ 
can be written as the flux of a monopole
$\Omega' =\nabla\times  \boldsymbol a' = \hat{\boldsymbol h'}/(2h'^2)$, 
through the Bloch sphere:
\begin{eqnarray}
C= \frac{1}{2\pi}\sum \int \boldsymbol \Omega' \cdot d\boldsymbol S_\bfk '\,.
\label{monopole}
\end{eqnarray}
Here, 
$\boldsymbol h'= \Big(  {\rm Re}\Delta(\bfk),    \rm Im\Delta(\bfk),  \xi_\pm \Big)$ are
the two monopoles' curvature fields.
The vector $\boldsymbol h'$  covers the Bloch sphere $C$ times. 
The north (N) and south (S) hemispheres
correspond to $h'_z=\xi$ being either positive or negative.
If the chemical potential, which is included in $h_0$, lies in the gap between 
the normal $\xi_\pm$ bands, then $C=0$ because $\boldsymbol h'$ 
stays always in one hemisphere. 
{\it Nonzero $C$ requires
the chemical potential to intercept 
at least
one of the normal bands} ($\xi_-$ or $\xi_+$)
and 
the N/S hemispheres are attained when $h'_z$ is outside or inside FS pockets.
{\it This  links $C$ to the topology of the FS.}
The other normal band does not contribute to $C$.
The N and S poles are attained at BZ points where 
$h'_x\pm ih'_y=\Delta^{(*)}(\bfk)$ vanishes. 
This motivates the choice of $p+ip$ pairing below.
A relevant example of a multiband superconductor
believed to have   $p+ip$ symmetry is Sr$_2$RuO$_4$
\cite{kallin,yada2014}

We recall that a theorem relating the topological indices of a 
superconductor to the FS topology
has been established by Sato\cite{sato2010} for the case 
where the normal bands have inversion symmetry and pairing
has odd parity. Our discussion above makes no requirement
on inversion symmetry or parity.
In the case of time-reversal invariant 
single band spin triplet superconductors, 
the topological indices were also shown to be related to
FS topology\cite{sato2009}.

For other choices of 
$\tau_j$ in equations   
(\ref{pairing})-(\ref{Huv1}) 
the direct product form
(\ref{uavb}) is no longer valid. 
We shall address those cases below 
by computing winding numbers for TRI momenta where $H$ is chiral.

\subsection{Model}

We write kinetic energy for $\up$-spin electrons as in equation (\ref{htau}),
where 
\begin{eqnarray}
h_x &=& \sin k_y \,, \qquad h_y =  -\sin k_x \,, \nonumber\\
h_z &=&  2t_1 \left(  \cos k_x + \cos k_y \right) + 4t_2 \cos k_x \cos k_y\,, \nonumber\\
h_0 &=& -\mu - t_1\left(  \cos k_x + \cos k_y \right) \,.
\label{zorro1}
\end{eqnarray}
We consider a spin triplet $p+ip$ pairing:
\begin{eqnarray}
\hat\Delta(\bfk) &=& d_z(\bfk) \sigma_1\otimes \tau_0 
\,,
\label{zorro2}
\end{eqnarray}
with $d_z(\bfk)=\Delta\left(  \sin k_x - i\sin k_y\right)$.
The pairing term (\ref{zorro2})  has odd parity under inversion.
This symmetry property reads\cite{Fu1}:
$
\tau_1 \hat\Delta(-\bfk) \tau_1
= - \hat\Delta(\bfk)$.

\begin{figure}
\begin{centering}
\includegraphics[width=3.2in]{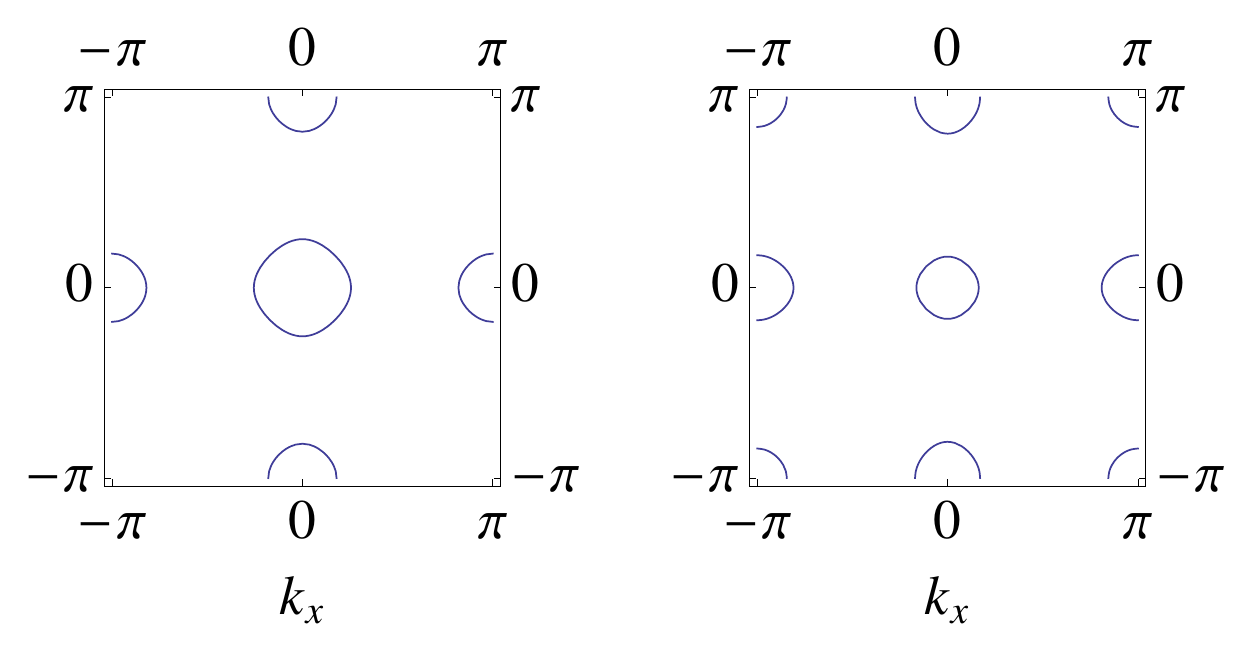} 
\caption{ \label{zorroFS3} Normal phase FS pockets for model   (\ref{zorro1})-(\ref{zorro2}),  with 
$t_2=-0.08$ (left, topological) and  $t_2=0.08$ (right, non-topological).}
\par\end{centering}
\end{figure}

\begin{figure}
\begin{centering}
\includegraphics[width=2.6in]{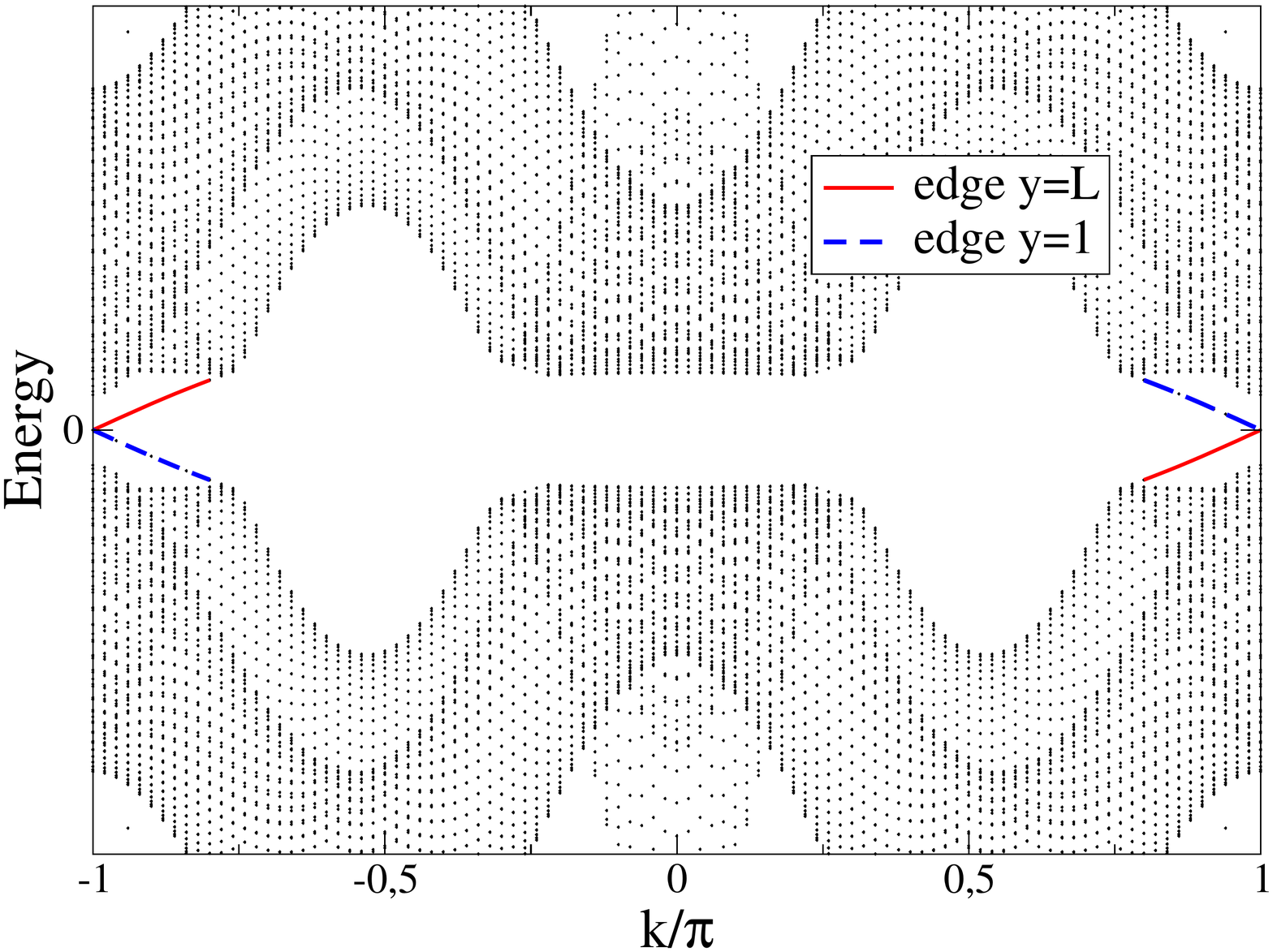}
\caption{ \label{zorromask} Spectrum 
for a ribbon geometry. $\Delta=0.1$,  $t_2=-0.08$.
The edge modes containing the MF are displayed in color.}
\par\end{centering}
\end{figure}

In the absence of Rashba or Dresselhaus spin-orbit couplings
the Bogolubov-deGennes matrix in the particle-hole basis,
$\left( (\hat c_\up) (\hat c_\dn) (\hat c_\up^\dagger) (\hat c_\dn^\dagger)\right)$,
splits into two  4x4 matrices. 
We
consider now only the subspace
 $\left( (\hat c_\up)  (\hat c_\dn^\dagger)\right)$.
The parameter choice
 $\mu=0.6$,
$t_1=0.07$, $t_2=-0.08$,
produces a FS with 3 pockets centered at  $(0,0)$, $(0,\pi)$, $(\pi,0)$
in the Brillouin Zone (BZ), 
as figure \ref{zorroFS3} (left panel) shows. 
We calculate the topological index (Chern number)  following 
 a method for multiband systems\cite{fukui}
and obtain  $C=+1$, indicating that one MF  exists.
 The energy spectrum for an infinite ribbon  in the yy (or xx) direction
is shown in figure \ref{zorromask}, where  
the MF  is clearly seen at longitudinal momentum  $\pi$.
If we consider the Andreev problem for a N/S interface along yy, 
we expect the MF to be detected when the incident electrons have transverse momentum $k_y=\pi$
and thus traverse only one FS pocket,  at $(0,\pi)$.

By reversing the sign of the hopping parameter $t_2$, a topologically trivial  phase is obtained,
with zero Chern number, and the FS now contains  4 pockets, as can be seen from the right
panel of figure \ref{zorroFS3}, 
and no MF's should exist.

One can also consider even parity pairing, 
which is realized by changing
the $\tau$ matrix in (\ref{zorro2}) to 
$\hat\Delta(\bfk) = d_z(\bfk) \sigma_1\otimes \tau_3$.
A direct computation of the ribbon spectrum and Chern number
yields the same topological indices as before. 
The ribbon spectrum remains gapped in the vicinity of  $k=\pi$, displaying 
the same edge mode, similar to that in Figure  \ref{zorromask}. 
This case lies outside the conditions of the 
theorem proved above. 

From equations (\ref{Huv1}) and  (\ref{zorro2})  
it is clear that the Hamiltonian for each subspace 
has the quiral properties\cite{ludwig}: 
\begin{eqnarray}
r_1 H(\bfk)r_1 &=& - H(\bfk) \qquad \mbox{if}\qquad k_x=0,\pi\,,
\label{chiralx}\\
r_2 H(\bfk)r_2 &=& - H(\bfk) \qquad\mbox{if}\qquad k_y=0,\pi\,.
\label{chiraly}
\end{eqnarray}
In either case one can 
rotate $H$ to the off-diagonal form,  
\begin{equation}
U^\dagger H(\bfk) U = \left( \begin{array}{cc}
0 & A \\
A^\dagger & 0
\end{array}\right)\,.
\end{equation}
The matrix $U$ is 
composed of the column eigenvectors of 
$r_1$ or  $r_2$, for each of the 
chiralities (\ref{chiralx})-(\ref{chiraly}),
yielding
\begin{eqnarray}
A_1(k_y)&=& - \Xi_\sigma +\hat\Delta\,,\\
A_2(k_x)&=&-\Xi_\sigma +i \hat\Delta\,,
\end{eqnarray}
respectively.
The phase of the determinant of $A_{1(2)}$
accumulates an amount $2\pi W_{1(2)}$, as its argument, $k_{y(x)}$, 
goes from $-\pi$ to $\pi$. The integer $W$ is 
the winding number.

For the choices $\tau_j=\tau_{0,3}$ in equation (\ref{pairing})
and $t_2=-0.08$ (topological case),
we obtain
$W_2(k_y=\pi)=1$ and 
$W_1(k_x=\pi)=-1$. If $t_2=+0.08$ (non-topological case), then 
all $W_{1,2}=0$.  For zero $k_x$ or $k_y$, we obtain $W_{1(2)}=0$. 
This agrees with the existence of a single MF mode with momentum $\pi$
along a ribbon, identified above.

From Figure \ref{zorroFS3} it is apparent that the FS pockets 
are approximately related through the nesting vector $\boldsymbol Q=(0,\pi)$
or  $(\pi,0)$, so one can ask about the effect of a charge (or spin) density wave (CDW
or SDW) on the topological properties. In the case of a SDW, we keep $s_z$ as a good quantum number.
The Chern numbers remain the same as before, the spectrum 
remains gapped and similar to that in Figure \ref{zorromask}. 
From the point of view of the winding numbers, however, there are quantitative changes
in the case of a CDW. If, for instance,   $\boldsymbol Q=(0,\pi)$ then the original BZ folds
such that $\pi/2<k_y<\pi/2$. 
For the topological case, $t_2=-0.08$, 
we obtain $W_1(k_x=\pi)=-2$ and $W_1(k_x=0)=0$. This means
that the MF still has momentum $\pi$ along a ribbon in the xx direction.
The winding number $W_2(k_y=0)=1$ because the points
$k_y=0,\pi$ are the same under the BZ folding. The MF has zero momentum for a ribbon
along $\boldsymbol Q$, because of the BZ folding.

In the case of a SDW, the Hamiltonian in equation (\ref{Huv1}) acquires
a term proportional to $r_0$ and no longer enjoys the chiral properties discussed.


\subsection{Induced pairing}

If pairing in the multiorbital superconductor is 
induced by a proximity effect, it would be desirable 
to know the possible matrices $\tau_j$ in equations 
(\ref{pairing}) and (\ref{Huv1}).
Such a way of engeneering a $p+ip$ supercondutor has been
discussed for single-band systems\cite{engeneering} and, 
more recently, for 
s-wave superconductors coupled to
Weyl semi-metals\cite{khanna}. 
A recent study\cite{balatsky} of the induced pairing symmetries
in a topological insulator's surface has been made,
in which pseudospin degree of freedom is taken into account.
It was found that a conventional local s-wave superconductor
can still induce spin triplet $p+ip$ pairing, due to  spin-momentum locking.    
In our model above there is no spin-momentum locking and induced
$p+ip$ from a conventional s-wave superconductor is not expected.
However, if a Rashba-type spin-orbit term is included in Hamiltonian
(\ref{XiDelta}), then spin-momentum locking will exist and 
the Chern number in equation (\ref{monopole}) 
will not be affected 
as long as the energy gap is not closed.

Suppose the Hamiltonian ${\cal H}(\Delta=0)$  of the  multiorbital
normal 2D system, in equations   (\ref{XiDelta}) and (\ref{zorro1}),
is coupled to a  single band superconductor  $\hat H_s$ 
through a tunneling term $\hat T$,
\begin{eqnarray}
\hat H_s &=& \sum_{\vk,\sigma} 
\left( \epsilon(\vk)-\mu \right)\hat b_{\vk,\sigma}^\dagger\hat b_{\vk,\sigma}
+ \Delta_{\sigma\sigma'}(\vk)  b_{\bfk,\sigma}^\dagger  b_{-\bfk,\sigma'}^\dagger
+  {\rm H.c.}\,, \nonumber\\ \label{hs}\\
\hat T &=& \sum_{\vk,j=1,2} V_{j} \hat c_{j,\vk\sigma}^\dagger b_{\vk,\sigma} 
+ {\rm H.c.}\,.
\end{eqnarray}
If the pairing term in equation (\ref{hs}) is a local spin singlet, the induced pairing
in the multiorbital layer contains 
the even parity local spin singlet terms (in $\sigma_2\tau_{0,1}$) as well as 
the $p+ip$ spin {\it singlet}, $d_z(\bfk)\sigma_2\tau_2$.
 The 
above spin triplet $p+ip$ term (\ref{zorro2}) is also induced. 
The latter is weaker,  however,
and the resulting system is not topological.

But 
 if the pairing term in equation (\ref{hs}) is a $p+ip$ spin triplet, the induced 
 triplet   term (\ref{zorro2})  is stronger than the local spin singlet term
($\sigma_2\tau_0$), and the multiorbital system has the topological properties
discussed above. 
An additional  {\it local} spin triplet  term (in $\sigma_1\tau_2$) is also induced.
Other smaller $p+ip$ spin triplet terms with odd (even) parity, 
in $d_z(\bfk)\sigma_1\tau_{1(3)}$, are also induced.
If one of the tunneling amplitudes is larger than the other
($V_1\gg V_2$) the induced spin singlet term is strongly supressed.

\section{The Andreev problem}

{The Chern number predicts the existence of Majorana edge states in the superconductor. 
Because a Majorana operator must be its own antiparticle, it must have  momentum either $0$ or $\pi$
along the edge.
In an Andreev scattering problem, the Majorana state should appear as an Andreev bound state
with transverse momentum ({\it i.e.} along the interface) $0$ or $\pi$.
 
In the framework of single band BTK theory, 
each Fermi pocket should contain one zero energy Andreev bound state. This is because the pairing function is odd, 
so that the electrons feel a sign change in the gap function upon specular reflection at the superconductor's
surface\cite{tanaka}. Thus 3 or 4 ABS's  or MF's are predicted by BTK theory, at conflict
with the topological properties.
The single MF observed when only a single FS pocket is crossed implies  that somehow
the two Andreev bound states predicted by BTK theory for the pockets $(0,0)$ and $(\pi,0)$ should
interfere destructively when the incident electron has transverse momentum $k_y=0$. 

We consider a N/S boundary along yy axis. 
In the framework of QWT\cite{waveguide,waveguide2}, 
the incident electron from a single band 
normal metal will split into the two pseudo-spin channels of the superconductor, as figure
\ref{tightexample} explains.
\begin{figure}[htb]
\centerline{\includegraphics[width=2.8in]{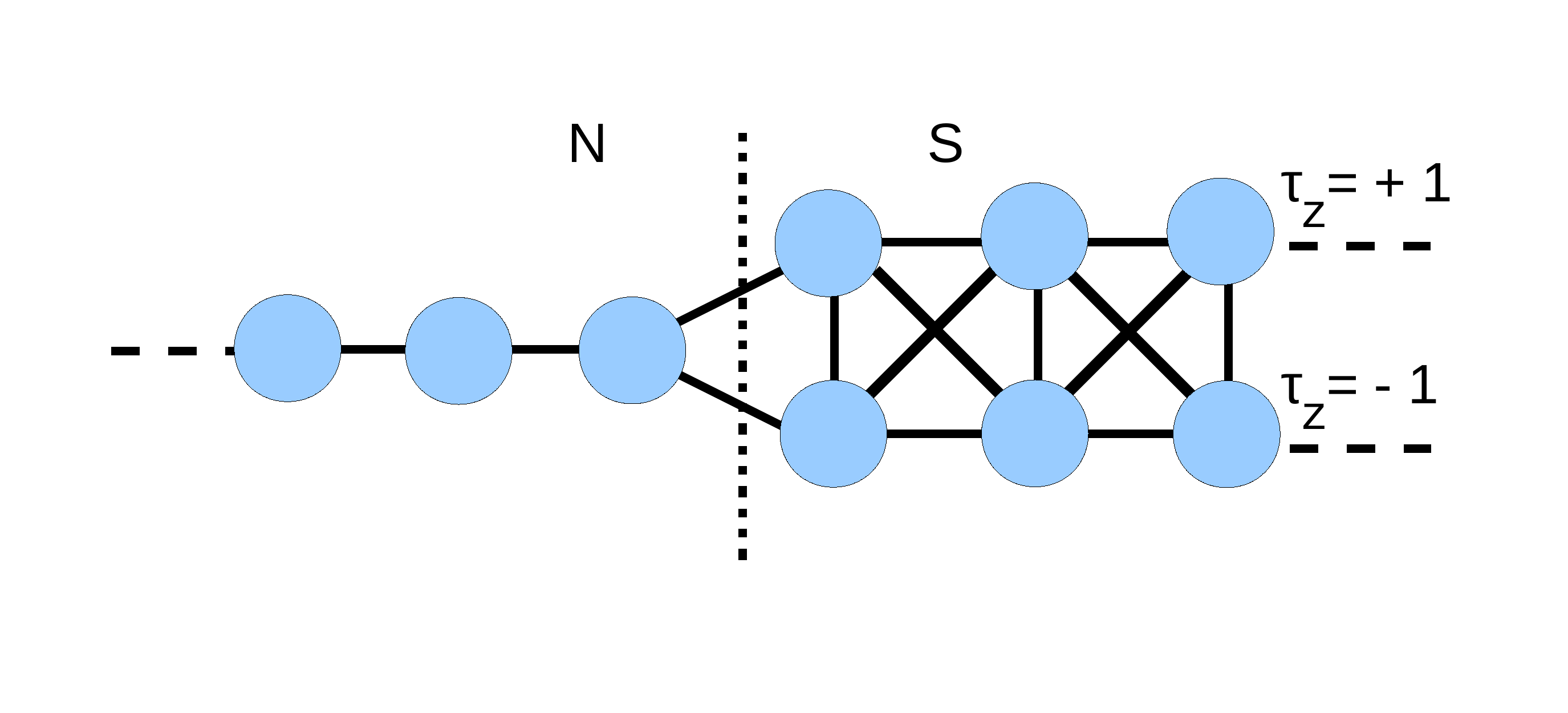}}
\caption{\label{tightexample} Tight-binding showing the splitting of the incoming electron
into two pseudo-spin channels of the superconductor, similar to a waveguide. 
}
\end{figure}
In the $N$ side, $x\leq 0 $, the wavefunction for electrons near the Fermi level 
is $\exp[i k_y y ]\psi_N$  where
\begin{equation}
\psi_N(x\leq 0)=\left(\begin{array}c 1\\0\end{array} \right)e^{ip_+x} + 
b\left(\begin{array}{c} 1\\0\end{array} \right)e^{-ip_+x} + 
a \left(\begin{array}{c} 0\\1\end{array} \right)e^{ip_-x}\,.
\label{inc}
\end{equation}
The momenta $p_\pm$ are close to the Fermi momentum $p_F$ and are fixed by the energy, E.
The amplitudes for electron reflection, $b$, and Andreev hole reflection, $a$,  allow us
to obtain the differential conductance as 
$
g_s= 1 + |a|^2 - |b|^2\,,
$
whereas the normal state conductance is just $g_n=1 - |b|^2$. We shall consider here only
$k_y=0$ or   $\pi$.
The transmitted waves into the superconductor are superpositions of 
wavevectors $\vk^\pm$, $\vq^\pm$ from two FS pockets
(see figure \ref{zorroFS3}) and the wavefunction for $x\geq 0$ is 
$\exp[i k_y y ]\psi_S$  where
\begin{eqnarray}
\psi_S(x\geq 0)&=& C\phi_{\vk^+}e^{ik^+x}+ 
D\phi_{\vk^-}e^{-ik^-x}\nonumber\\  &+& 
E\phi_{\vq^+}e^{iq^+x}+ 
F\phi_{\vq^-} e^{-iq^-x}\,,
\label{trans}
\end{eqnarray}
where each $\phi_{\vk}$ denotes a
four-dimensional column eigenvector of the BdG matrix,  $H$.
The x-components of the momenta,
 $k^\pm$,   $q^\pm$, 
 are chosen so that the group velocity is positive for energy E above the gap.
 For subgap energies, the momenta have a positive imaginary part.

In QWT the matching conditions for wave functions (\ref{inc}) and (\ref{trans}) at $x=0$ 
are written as\cite{waveguide,waveguide2} 
\begin{eqnarray}
\psi_N(0) \otimes   \left(\begin{array}{c}
1  \\ 1 \end{array} \right) &=& \psi_S(0)\,,
\label{continuidade}\\
\partial_{k_x} \hat H_N \psi_N(0) 
&=& \left(\begin{array}{cccc}
1 & 1 & 0 & 0 \\ 0 &0 & 1& 1 \end{array} \right) \cdot 
\partial_{k_x} \hat H_S
\psi_S(x=0)\,.\nonumber\\ 
\label{correntes}
\end{eqnarray}
Here, $H_N$ denotes a BdG Hamiltonian matrix for the normal metal.  
Interface disorder can be accounted for\cite{waveguide} 
by making  the replacement:  
$1-b \rightarrow 1-b-2iZ(1+b)p_F/p_+$ 
and
$a \rightarrow a(1-2iZp_F/p_+)$
in the right-hand side of equation (\ref{correntes}), 
and  where $Z$ denotes the 
BTK parameter\cite{btk}.

 \section{Results}
  
According to QWT\cite{waveguide}, 
 the condition for the existence of ABS 
is obtained from the  4x4 matrix $\Lambda$ 
composed of  the four column
vectors $\phi_{\vk^+}$,   $\phi_{\vk^-}$,
 $\phi_{\vq^+}$,  $\phi_{\vq^-}$,  in equation (\ref{trans}).
 The condition then reads
 \begin{equation}
{\rm Det\ }\Lambda = 0\,.
\label{detlam}
\end{equation}
We checked that  condition (\ref{detlam}) does not hold 
for transverse momentum $k_y=0$ either in the 
topological ($t_2=-0.08$) or in the  non-topological case ($t_2=+0.08$).
For transverse momenta $k_y=\pi$,
 equation   (\ref{detlam})  is verified only in the topological case.  
This means that the quantum interference effects from the two FS pockets effectively 
anihilate the two ABS's that  would be predicted by single pocket BTK theory. 

One might be tempted to read  condition (\ref{detlam}) as the requirement that the
linear system in equation (\ref{continuidade})  be homogeneous and the 
wave function  $\psi_S$  be made to vanish\cite{tanaka2012}  at the N/S boundary. 
This would be  incorrect, however, as it would imply the conductance $g_s$ to vanish. 
On the contrary, the conductance $g_s$ is finite and  independent
of the disorder parameter $Z$ at the energy value where equation (\ref{detlam}) is 
obeyed\cite{waveguide,btk,tanaka}.

For the non-topological case, the  differential conductance at normal incidence,
 $k_y=0$, is shown in figure \ref{gs+008py0} (left)
 as a function of energy (which in an actual experiment is obtained from
 the voltage bias). It is seen that the quantum interference
supresses quasi-particle transmission, $g_s$,  
as $E\rightarrow 0$ and the effect is even more pronounced as disorder increases.
A similar result is obtained at   $k_y=\pi$.
Features such as peaks and dips 
are visible when the energy $E$ crosses the superconducting gaps on the Fermi pockets.
   
For the topological case, the differential conductance at normal incidence, $k_y=0$, is shown in 
figure \ref{gs-008py0}. 
A similar destructive interference is observed at low energy.

For transverse momentum $k_y=\pi$, the ABS (MF) leaves its imprint on the conductance,
as figure \ref{majorana} shows.  The differential conductance attains the maximum value
$g_s=2$ at $E=0$, independent of Z.

\begin{figure}
\begin{centering}
\includegraphics[width=1.6in]{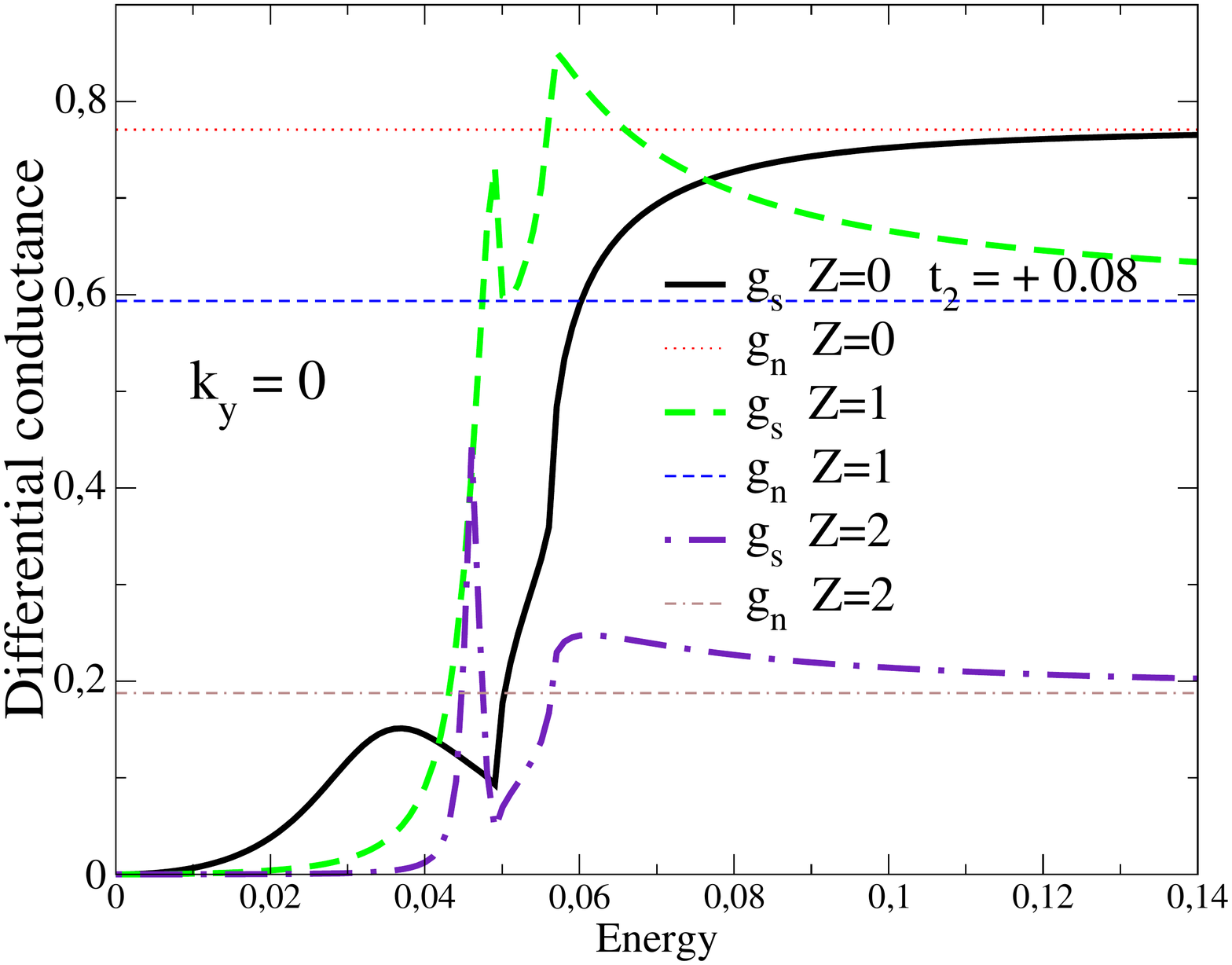}
\includegraphics[width=1.6in]{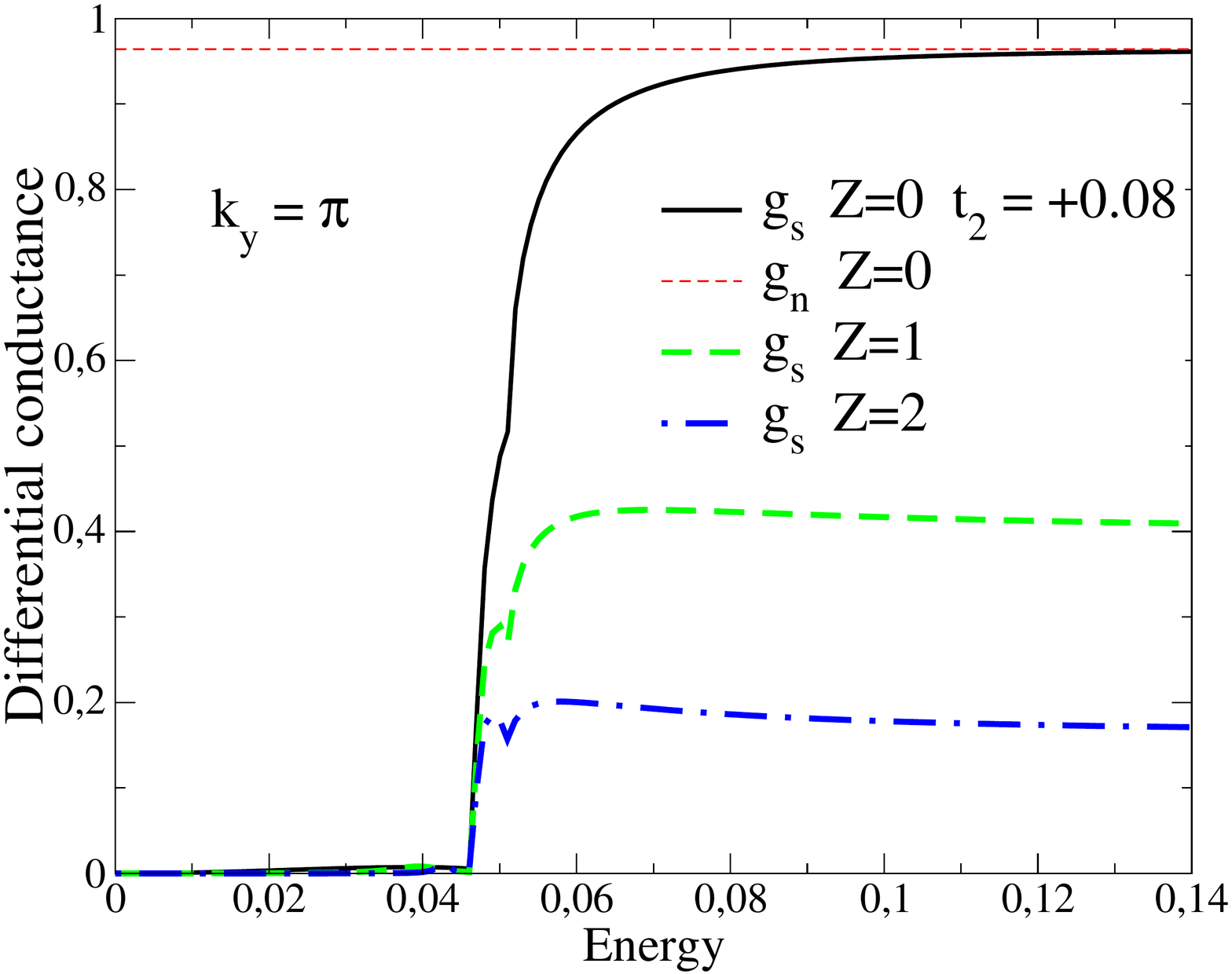} 
\caption{ \label{gs+008py0} Differential conductance at normal incidence (left) 
and for $k_y=\pi$ (right). 
Model (\ref{zorro1}) -(\ref{zorro2})
with $\Delta=0.1$
and $t_2=0.08$ (non-topological). 
Dashed line: normal state differential conductance.}
\par\end{centering}
\end{figure} 
\begin{figure}
\begin{centering}
\includegraphics[width=2.6in]{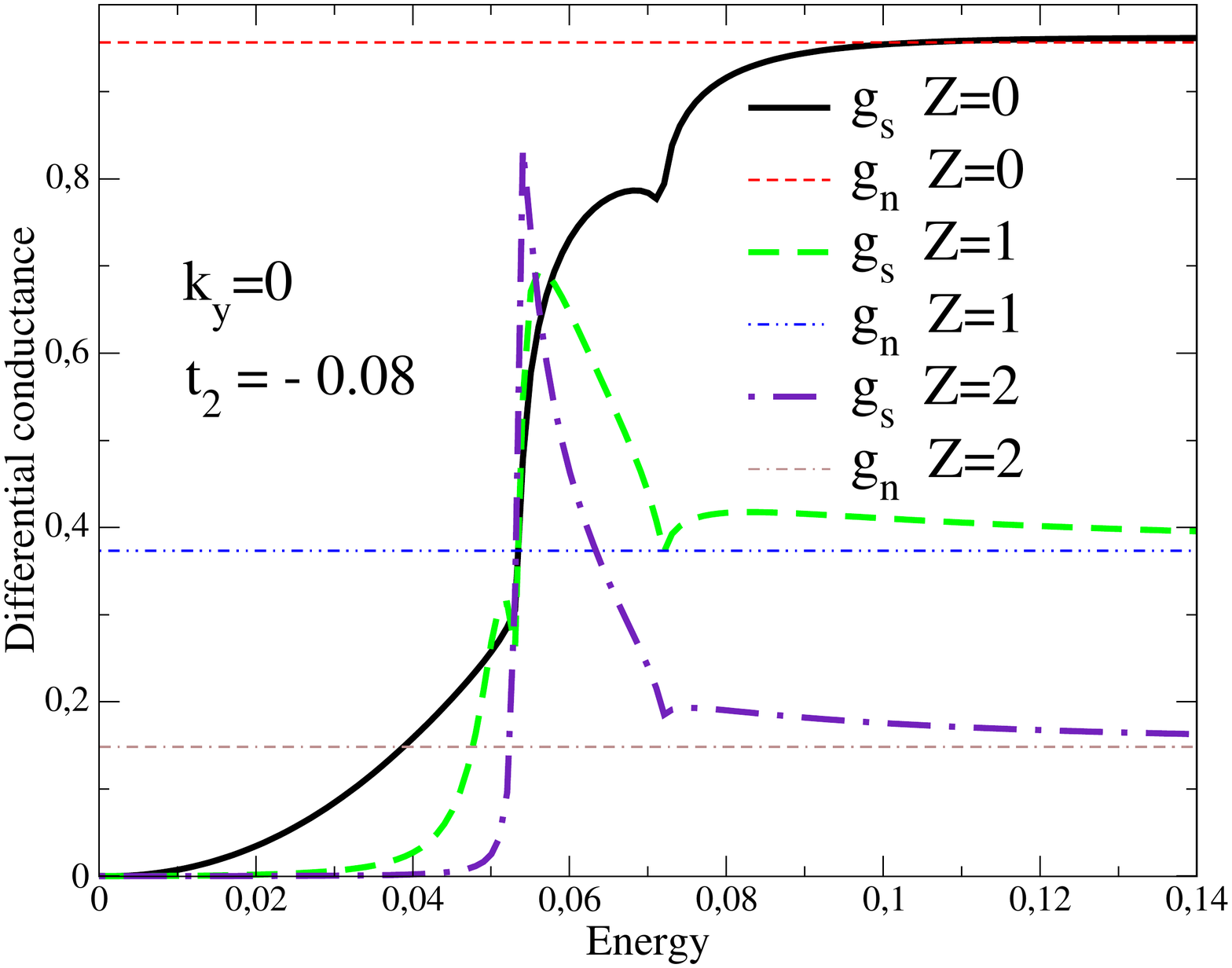} 
\caption{ \label{gs-008py0} Differential conductance at normal incidence,  for 
the same model as in Figure 
\ref{gs+008py0} with  $t_2=-0.08$ (topological).
Dashed line: normal state differential conductance.}
\par\end{centering}
\end{figure} 
\begin{figure}
\begin{centering}
\includegraphics[width=2.4in]{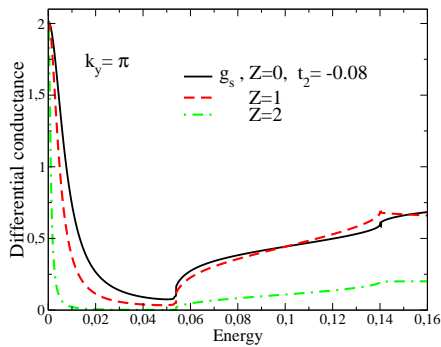}
\caption{ \label{majorana} Differential conductance for
 the topological case, showing the signiture of the MF.}
\par\end{centering}
\end{figure} 

\section{Summary and Conclusions}

We have studied a model of a multiorbital 2D superconductor which is realized by 
introducing $p+ip$ Cooper pairing, diagonal in the pseudospin channel,  in an otherwise TI. 
Under these conditions, the topological properties of the TI were shown to become irrelevant.
The topological properties of the multiorbital  superconductor depend on the FS in the normal phase. 
 If the  FS contains several sheets, we have shown that 
when the number of Fermi pockets is odd, the MF  is in the single pocket that is traversed
by the quasi-particles, while 
the other pair of  pockets interfere destructively. 
This result from QWT  
reconciles the number of MFs
with the Chern number, in contrast to BTK theory.
In addition to the destruction of the MFs,  
the waveguide interference effects also produce a vanishing conductance at $E=0$, 
when a pair of FS pockets
is traversed by the quasi-particles.

\section{Acknowledgments}

We acknowledge the hospitality of CSRC, Beijing, China,
where part of this work has been carried out.
We would like to thank financial support from 
Funda\c{c}\~ao para a Ci\^encia e Tecnologia 
(Project EXPL/FIS-NAN/1728/2013 and Grant No. PEST- OE/FIS/UI0091/2011).


\end{document}